\documentclass[aps,prl,amsmath,amssymb,longbibliography,noeprint,10pt,twocolumn]{revtex4-2}
\usepackage{hyperref}
\usepackage{amsmath}
\usepackage{graphicx}
\usepackage[utf8]{inputenc}
\usepackage{amsthm}
\usepackage{relsize}
\usepackage{aas_macros}
\usepackage{bbm}
\usepackage{color}
\usepackage{longtable}
\usepackage{enumitem}
\usepackage{multirow}
\usepackage{tikz}
\usepackage{changepage}
\usepackage{orcidlink}
\usepackage{multirow}
\usepackage{dcolumn}
\usepackage{xspace}

\def\be{\begin{equation}}
\def\ee{\end{equation}}

\newcommand{\eq}[1]{(\ref{#1})}

\newcommand{\bM}{{\mathbf{M}}} 
\newcommand{\bN}{{\mathbf{N}}} 
\newcommand{\dd}{{\rm{d}}}
\newcommand{\HMP}{\mbox{$20{\kern 0.0833em}480$ }}
\newcommand{\bQ}{{\mathbf{Q}}}
\newcommand{\HD}{{\rm{HD}}}
\newcommand{\CURN}{{\rm{CURN}}}
\newcommand{\ke}{{\kern 0.0833em}}
\newcommand{\BF}{{\mathrm{BF}}}
\newcommand{\efac}{{\mathrm{EFAC}}}
\newcommand{\equad}{{\mathrm{EQUAD}}}
\newcommand{\ecorr}{{\mathrm{ECORR}}}
\newcommand{\ENT}{\texttt{Enterprise}\xspace}
\newcommand{\TEM}{\texttt{Tempo2}\xspace}
\newcommand{\PINT}{\texttt{PINT}\xspace}
\newcommand{\DISC}{\texttt{Discovery}\xspace}
\newcommand{\PAR}{\texttt{.par}\xspace}
\newcommand{\TIM}{\texttt{.tim}\xspace}
\newcommand{\FEA}{\texttt{Feather}\xspace}

\newcounter{emitem}
\renewcommand{\theemitem}{EM\arabic{emitem}}

\begin{document}

\title{\texorpdfstring{Stochastic gravitational-wave background search\\ 
  using data from five pulsar timing arrays}{sgwbsudffpta}}

\date{\today}

\author{Wang-Wei Yu \orcidlink{0000-0002-0844-7080}}
\email{wangwei.yu@aei.mpg.de}
\affiliation{Max Planck
  Institute for Gravitational Physics (Albert Einstein Institute),
  Leibniz Universit\"at Hannover, Callinstrasse 38, D-30167, Hannover,
  Germany}

\author{Bruce Allen \orcidlink{0000-0003-4285-6256}}
\email{bruce.allen@aei.mpg.de}
\affiliation{Max Planck Institute for
  Gravitational Physics (Albert Einstein Institute), Leibniz
  Universit\"at Hannover, Callinstrasse 38, D-30167, Hannover,
  Germany}
 
\begin{abstract}
  \noindent
  Using public pulse time-of-arrival data from five pulsar timing arrays (PTAs), we search for a stationary, isotropic, and unpolarized nHz stochastic gravitational-wave background (SGWB). This analysis is more sensitive than previous individual PTA searches because the combined 121-pulsar dataset is about four times larger than any single PTA's. For pulsars observed by multiple PTAs, we employ a new ``direct combination'' method to merge their astrophysical models and data. This avoids the challenge of reconciling different PTA timing models to obtain a single ``best'' model.  A central result of our analysis is the posterior distribution of the amplitude $A_{gw}$ and exponent $\gamma_{gw}$ of the putative SGWB energy-density spectrum, modeled as a power law in frequency.  While these results are consistent with a nonzero SGWB amplitude $A_{gw}$, the statistical significance-assessed via a Bayesian odds ratio and noise-marginalized false-alarm probabilities ($p$-values) for three detection statistics-remains below the conventional $5\sigma$ threshold for a confident detection.  The inter-pulsar timing-residual correlation, reconstructed as a function of angle $\theta$ between the pulsar lines of sight, matches the Hellings and Downs (HD) prediction.
\end{abstract}


\maketitle 

\emph{Introduction} -- Pulsar timing arrays (PTAs) aim to detect
low-frequency gravitational waves (GWs) by measuring the effects
these waves induce on the arrival times of pulsar pulses. These
GWs create ``timing residuals'' which are correlated between
pulsars~\cite{taylor2021nanohertz, FAQ}.  Decades-long datasets with
submicrosecond timing precision yield GW amplitude (strain)
sensitivities of order $A_{gw} \sim 10^{-15}$ at nHz frequencies.

Supermassive black hole binaries (SMBHBs) are one source of
GWs. If the Universe contains many of them, the incoherent
superposition of their GW emissions would create a stochastic GW
background (SGWB)~\cite{agazie2023nanogravastrophysics,burke2018astrophysics,phinney2001practical}
with a power-law spectrum in the nHz band.

Different PTA collaborations have reported varying levels of evidence
for a nHz SGWB. The Parkes PTA (PPTA~\cite{PPTA_new}) observed 30
pulsars over 18~yr, concluding that the data show ``no support for or
against''.  The European and Indian PTAs (EPTA and
InPTA~\cite{EPTA_new23}) observed 25 pulsars and reported ``marginal
evidence'' in the 25-yr dataset and ``evidence'' when only the last
10~yr were analyzed. The Chinese PTA (CPTA~\cite{CPTA_new}) observed
57 pulsars over 3~yr, and reported ``some evidence''.  The North
American Nanohertz Observatory for Gravitational Waves
(NG~\cite{NANOGrav_new}) observed 67 pulsars over 15~yr and reported
``compelling evidence''.  Most recently, the MeerKAT PTA
(MPTA~\cite{miles2025meerkat}) observed 83 pulsars over 4.5~yr, and,
depending on assumptions about pulsar noise processes, reported either
``high significance'' or ``no significance'' (see
also~\cite{RutgerWarning}).  None reach the conventional $5\sigma$
significance threshold required to claim a
detection~\cite{IPTAchecklist}.

\begin{figure}[h]
\centering
\includegraphics[width=0.48\textwidth]{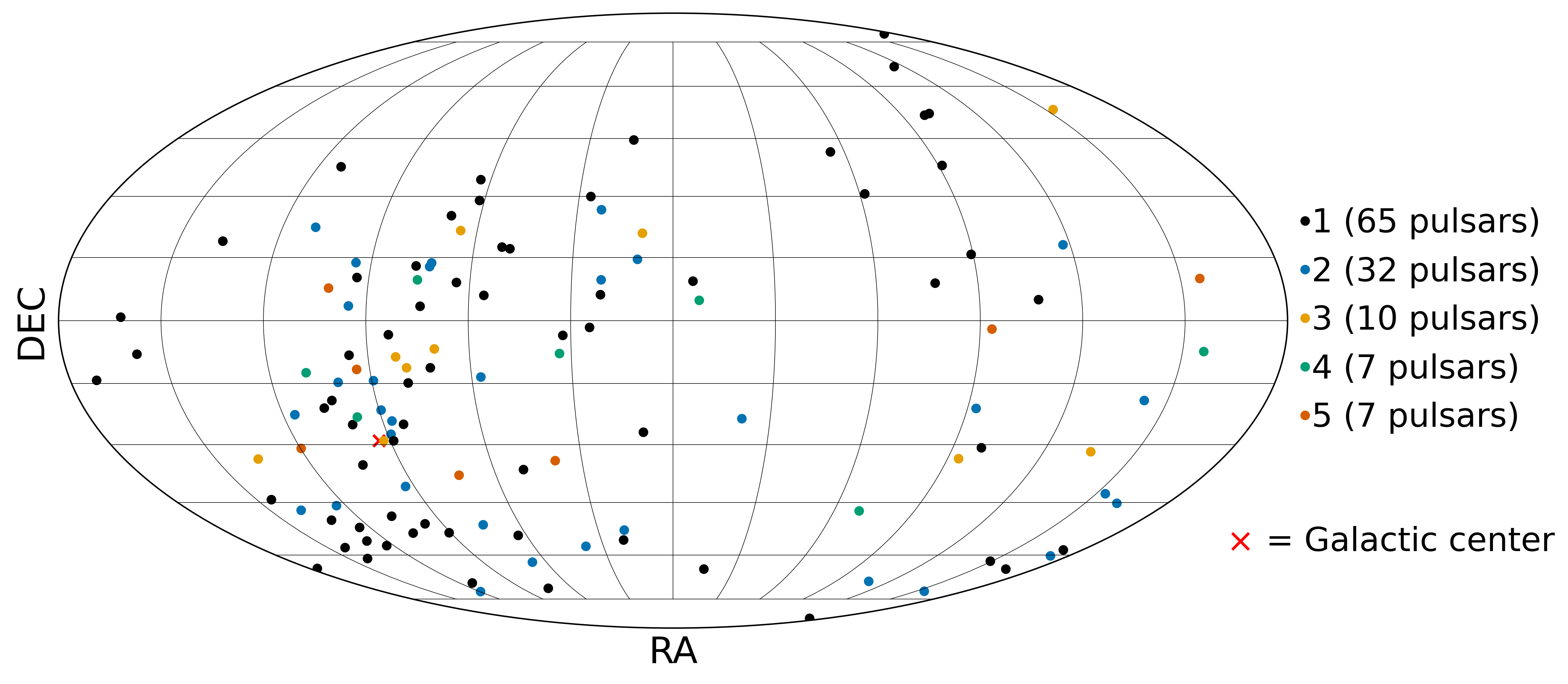}
\vskip -0.1in
\caption{\label{fig:iptapulsars} A Mollweide projection showing
  sky locations of the 121 pulsars included in our analysis.  Colors
  indicate how many PTAs provided data for that pulsar.}
\end{figure}

All of these PTAs (apart from CPTA) have released their data.  Using
this full public dataset increases the total number of distinct
pulsars to $N_p=121$ as shown in Fig.~\ref{fig:iptapulsars}.  PTAs
search for an SGWB by correlating pairs of pulsars, whose number
scales as $N_p^2$.  Furthermore, since many pulsars are observed by more
than one collaboration (on average, 1.8) there are also more data
samples per pulsar.  Hence, analysis of the combined dataset should
significantly increase the GW strain sensitivity and the detection
confidence~\cite{goldstein2018null,perera2019international,
  baier2024tuning,FAQ}.

We analyze this five PTA dataset~\cite{EPTAdataSource,
  InPTAdataSource, MPTAdataSource, NANOGravdataSource, PPTAdataSource}
using a recently proposed ``direct combination'' method~\cite{DynComb,
  DynCombCode}.  This allows us to combine the data with minor
modifications, as described below. Tests of this
method~\cite{DynComb}, carried out with the International Pulsar
Timing Array~\cite{hobbs2010international} (IPTA) Data Release 2 (DR2)
combined dataset from 2019~\cite{perera2019international}, suggest
that it produces results identical to traditional (but more
time-consuming) data combination techniques.

Aside from this, our analysis is standard. We use the same tools and
methods as the individual PTAs, as implemented in
{\ENT}~\cite{ellis2019enterprise} and in the next-generation PTA
data-analysis package {\DISC}~\cite{nanogravdiscovery}.  Both use
parameterized likelihood models and Markov chain Monte Carlo (MCMC) to
generate posterior samples (``posteriors'') approximating the
parameters' posterior distributions.

Note: additional details are provided in End Matter (EM) and
supporting tables and plots in the Supplementary Material (SM).

\emph{Methods, Models, and Conventions} -- Because we use standard
methods, this paper is brief.  When we follow the PTA analyses
described above~\cite{PPTA_new, EPTA_new23, CPTA_new, NANOGrav_new,
  miles2025meerkat}, we give only a cursory description and cite the
relevant literature. If we do something different, then more details
are provided.

As is usual in this field, our analysis considers two alternative
hypotheses or ``models'':
\begin{adjustwidth}{0.5em}{0.5em}
-- The \emph{signal hypothesis}, denoted by ``HD'', is a model with
two contributions to the pulsar pulse arrival-time residuals.  The
first is pulsar noise, assumed to be (a) uncorrelated between pulsars
and (b) a sum of stationary ``red'' (RN) and ``dispersion measure'' (DM)
noise plus a nonstationary ``white noise'' (WN) component.  The second
contribution arises from a stationary, unpolarized and isotropic SGWB,
as described by the general theory of relativity (GR). This SGWB
component, determined by parameters $A_{gw}$ and $\gamma_{gw}$ defined
below, induces timing-residual correlations between pulsar pairs,
which are assumed to follow the Hellings and Downs (HD)
prediction~\cite{1989NASCP3046...93H,allen2023variance,FAQ} given
in~\eq{e:HDmat}.  Both contributions are assumed to arise from (the
sum of) many independent sources; by the central limit theorem, the
resulting timing residuals form a zero-mean Gaussian process, which is
fully characterized by its covariance matrix $\bf C$.
\\
-- The \emph{null hypothesis} is called ``common uncorrelated red
noise'' and is denoted by CURN.  Again, the timing residuals are
assumed to be drawn from a zero-mean Gaussian process, but this time
entirely characterized by the covariance matrix ${\bf C_0} = {\rm
  diag} \, \bf C$. This zeros all elements where the row and column have
different pulsars, see~\ref{EM:1}. Since our null hypothesis
\emph{is defined via our signal hypothesis}, it has \emph{exactly} the
same parameters/noise-models as the signal hypothesis.
\end{adjustwidth}
Using {\DISC}~\cite{nanogravdiscovery} as a tool, and
conditioning the Gaussian likelihood on the data and the priors, we
generate posteriors for the HD and CURN models.  These are sets of
inferred parameter values most consistent with the data and priors,
thereby characterizing the posterior distributions of the model
parameters.

The most important model parameters are $A_{gw}$ and $\gamma_{gw}$,
which control the power-law amplitude and exponent for the SGWB power
spectrum.  To define these, let $\dd \rho_{gw}$ denote the SGWB energy
density in the frequency range $[f,f+ \dd f]$ and $\rho_{cr} = 3
c^2 H_0^2/8 \pi G \simeq 7.7\times10^{-10}\ \mathrm{J}/\mathrm{m}^3 $
denote the critical (closure) energy density of the Universe. Here,
$c$ is the speed of light, $G$ is Newton's gravitational constant, and
$H_0 \simeq 67.4~\mathrm{km}/\mathrm{s \, Mpc} \simeq 2.2 \times
10^{-18}\, \mathrm{Hz}$ is the current Hubble expansion rate.  Then
\begin{equation}
  \label{e:OmegaOfF}
  \Omega_{gw}(f) \equiv \frac{f}{\rho_{cr}} \frac{\dd \rho_{gw}}{\dd f} =
  \frac{2 \pi^2}{3H_0^2} A^2_{gw} f_r^2 \Bigl(\frac{f}{f_r} \Bigr)^{-\gamma_{gw} + 5} \, ,
\end{equation}
where $f_r = 1/\mathrm{yr} \simeq 31.7\, \mathrm{nHz}$ is a standard
reference frequency.  Together with our previous assumptions, this
spectral model fully characterizes the SGWB over the 14 frequency
bins of our analysis $ 2 \lesssim f/\mathrm{nHz} \lesssim 30$.

One result of this analysis is posterior probability distributions for
$A_{gw}$ and $\gamma_{gw}$.  We consider models in which $\gamma_{gw}$
is free, and those where it is constrained to $\gamma_{gw}=13/3$.  The
latter would be expected from a population of SMBHBs in circular
orbits, provided that GWs dominate the energy loss at nHz frequencies.

One way to assess detection confidence is via frequentist detection
statistics~\cite{BestStatistic}.  We consider three different
statistics $D$ used within the PTA community.  Each $D$ is a quadratic
function of the timing residuals, with weights determined by $\bf C$
and $\bf C_0$.  Because each posterior sample corresponds to a
different noise model and SGWB---and thus different covariances---the resulting
statistic values and associated false-alarm probabilities ($p$-values)
vary across samples. We therefore characterize detection confidence
using the posterior distribution of these $p$-values, whose mean
corresponds to the posterior predictive
$p$-value~\cite{vallisneri2023posterior}, analogous to the
noise-marginalized optimal statistic~\cite{vigeland2018noise}.

Some past work uses ad hoc techniques to estimate the false-alarm
probability $p$, see~\ref{EM:2}.  We compute $p$-values using a
statistically rigorous analytic approach~\cite{hazboun2023analytic,
  NG3.2sigma, RutgerPvalues, BestStatistic}. Since each statistic $D$
is a quadratic form in the data, its null (CURN) distribution is a
generalized $\chi^2$; the probability of exceeding the observed value
is obtained analytically for each posterior sample.

To aid interpretation, we also report the $p$-values as ``equivalent
$q\sigma$'' levels: $q>0$ is chosen so that a random draw from a
zero-mean Gaussian with variance $\sigma^2$ has probability $p$ of
exceeding $q\sigma$.  This ``one-sided $p$ value'' satisfies $2p =
\mathrm{erfc}(q/\sqrt{2})$. 
For example, $1\sigma$ corresponds to $p =
16\%$, $3\sigma$ corresponds to
$p = 0.14\%$, and
$5\sigma$ corresponds to $p = 2.9\times10^{-7}$.
The latter, a
false-alarm probability of less than one per three million, is the
traditional threshold for confident detection~\cite{IPTAchecklist}.

\emph{Analysis of the five PTA dataset} -- The data from the five PTAs
come from 121 millisecond pulsars, as shown in
Fig.~\ref{fig:iptapulsars}.  These consist of the second EPTA data
release (10-yr subset)~\cite{EPTA_data,EPTAdataSource}, the first
InPTA data release~\cite{InPTA_data,InPTAdataSource}, the MPTA 4.5-yr
data release~\cite{MPTA_data,MPTAdataSource}, the NG 15-yr data
release~\cite{agazie2023nanograv,NANOGravdataSource}, and the third
PPTA data release~\cite{PPTA_data,PPTAdataSource}. Each PTA's public
open dataset has one {\PAR} file and one or more {\TIM} files per
pulsar.

The data we analyze consist of $N= 1 \ke 090 \ke 206$ times of arrival
(TOAs): the complete set of observations from 25 EPTA pulsars, 14
InPTA pulsars, 83 MPTA pulsars, 68 NG pulsars, and 32 PPTA pulsars.
We merge these using the direct combination method~\cite{DynComb,
  DynCombCode}. For each pulsar, this adopts one set of parameter
values $\beta_0$, defining its nominal astrophysical
model~\cite[Eq.~7.2]{taylor2021nanohertz}; for details see~\ref{EM:3}
and SM Sec.~\ref{s:workflow}.  It works because each pulsar has exact but
unknown values $\beta$ for these parameters (for example, for the sky position).
They enter the analysis via the nominal value
$\beta_0$~\cite[Eq.~7.2]{taylor2021nanohertz} and via a ``design
matrix'' $\bM$~\cite[Eq.~7.4]{taylor2021nanohertz}, which is the rate
of change of the pulse arrival-time with respect to variations in
$\beta$, evaluated at $\beta_0$. In the (tiny) region of interest,
$\bM$ varies slowly with those parameters, so computing $\bM$ at the
same nominal $\beta_0$ for a given pulsar introduces negligible
errors. The subsequent marginalization over $\epsilon = \beta -
\beta_0$~\cite[Eq.~7.6]{taylor2021nanohertz} fully accounts for the
uncertainties in $\beta$ and for the variations in its nominal value
across PTAs.  Similarly, the RN and DM noise are intrinsic to the
pulsar and to the interstellar medium (ISM), so must be the same
across PTAs.  We therefore adopt the physically motivated DMGP
model~\cite{DMGP}, used by all PTAs except NG and InPTA, which use the
DMX model.  The design matrix $\bM$ allows the effects of DM
fluctuations on the timing residuals to be computed from the
fixed-in-time DM timing residuals.

The next analysis step is single-pulsar noise modeling, which
determines each pulsar's WN covariance matrix $\bN$ appearing in the
Gaussian likelihood~\cite[Eq.~7.27]{taylor2021nanohertz}
(see~\ref{EM:4}). This modeling yields two or three parameters per
pulsar and backend-$\efac$, $\equad$, and (in some cases) $\ecorr$.
NG, PPTA, MPTA, and seven EPTA backends use all three
parameters~\cite[Eq.~2]{agazie2023nanograv}, while InPTA and the
remaining EPTA backends use only $\efac$ and
$\equad$~\cite{goncharov2025reading}.

\begin{table}[h]
  \begin{tabular}{lll}
    Parameters & Uniform on range  $\quad$ & Stage \\
    \hline
    EFAC               $\quad$ & (0.01, 10)   $\quad$& S\\
    $\log_{10}$ EQUAD  $\quad$ & (-8.5, -5)   $\quad$&  S\\
    $\log_{10}$ ECORR  $\quad$ & (-8.5, -5)   $\quad$&  S\\
    $\gamma$           $\quad$ & (1,7)        $\quad$&  S, F\\
    $\log_{10} A $     $\quad$ & (-20, -11)   $\quad$&  S, F\\[0.4ex] 
        \hline
  \end{tabular}
  \caption{\label{tab:priors} Priors for the WN parameters, and for
    the amplitudes/exponents of the power-law RN, DM and SGWB noise
    spectra. ``S'' denotes single pulsar analysis; ``F'' denotes
    multipulsar analysis. The WN dictionary found in the ``S'' stage is
  used in the ``F'' stage.}
\end{table}

The WN parameters for each pulsar are inferred from all of its
\TIM-file TOA data using \texttt{JAXopt}~\cite{jaxopt}, with priors
from Table~\ref{tab:priors} (rows marked ``S''); the amplitude/exponent
priors are applied to all power laws. For $\gamma<7$, the $f\to0$
divergences are absorbed by the timing model, while for $\gamma>1$ the
$f\to\infty$ divergences are absent~\cite{MarcoEtAl,Legendre}. The
parameter $\equad$ is expressed in the
\texttt{Tempo}/\TEM/\PINT\ convention
(\texttt{T2EQUAD})~\cite{TEMPO,Hobbs:2006cd,arXiv:2012.00074}. The
output is a ``WN dictionary'' of best-fit parameters, from which $\bN$
is then determined using the \TIM file contents.

The final stage is multi-pulsar analysis, following the procedures
described in~\cite{taylor2021nanohertz,NANOGrav_new}.  We use both
{\ENT}~\cite{ellis2019enterprise} and the next generation PTA
data-analysis package, {\DISC}, which is optimized via JAX for
graphics processing units~\cite{jax2018github}.  The use of {\DISC}
has significantly sped up our analysis.

Within the {\DISC} framework, we use the WN dictionary 
plus
the SGWB priors listed in
Tab.~\ref{tab:priors} with an ``F'' in the final column. As described
above, we consider two priors for the SGWB power-law power-spectrum
exponent $\gamma_{gw}$.  One is the uniform range given in
Tab.~\ref{tab:priors}, the other is fixed to $\gamma_{gw} = 13/3$.
Our analysis uses the
 {\DISC}~\cite{nanogravdiscovery}
default
Hamiltonian No-U-Turn Sampler
(NUTS)~\cite{neal2011mcmc,hoffman2014no}, as implemented in
NumPyro~\cite{phan2019composable,bingham2019pyro,betancourt2017conceptual},
see~\ref{EM:5}.

We generated \HMP posterior samples under the HD signal hypothesis,
and the same number under the CURN hypothesis.  Each gives the
amplitudes and exponents of the power-law models describing the pulsar
(red and DM) noise, and the SGWB. We use these to evaluate the
frequentist detection statistics and their corresponding false-alarm
probabilities ($p$-values) for each posterior sample.

\begin{figure}[h]
\centering
\includegraphics[width=0.45\textwidth]{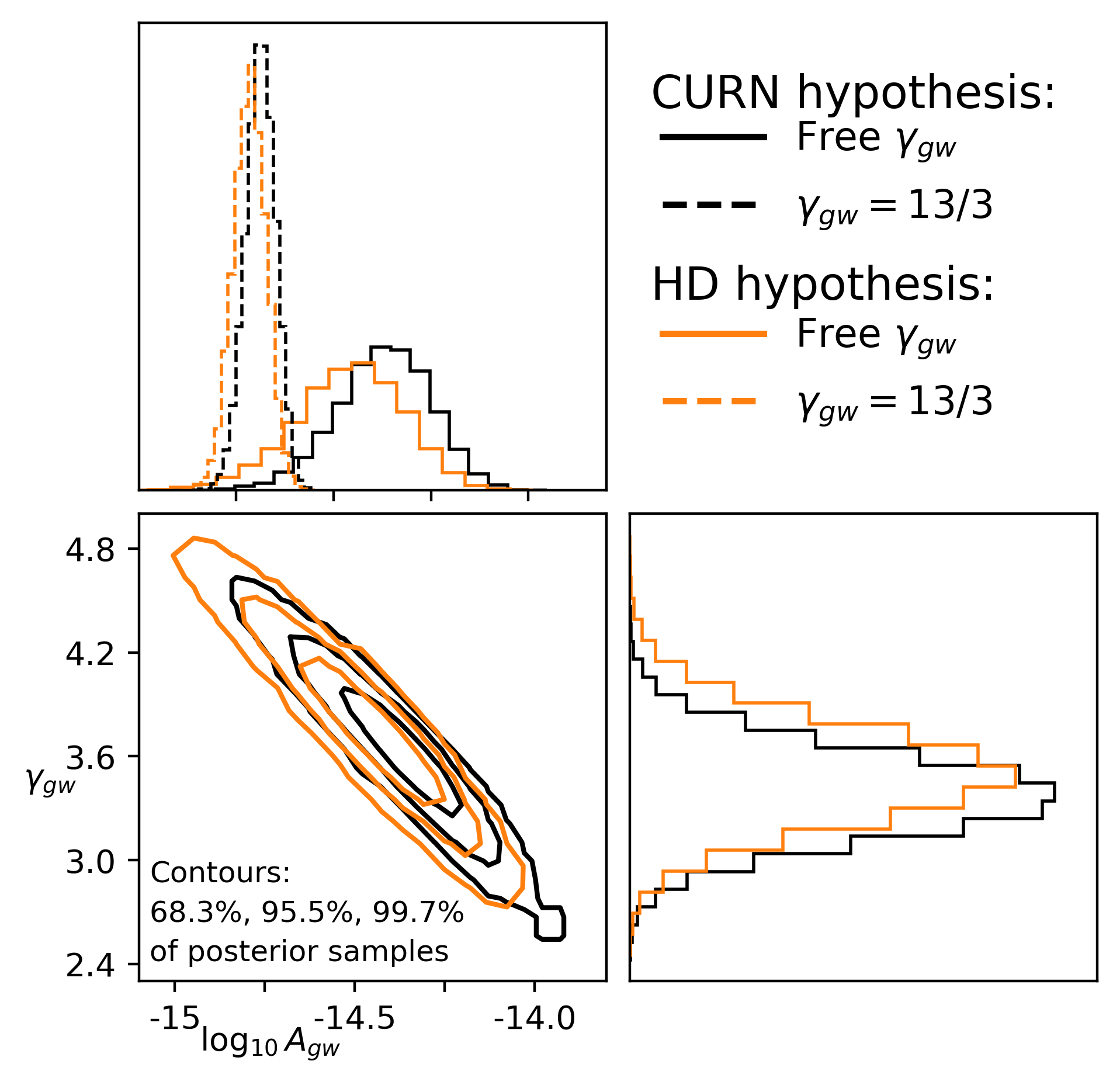}
\vskip -0.2in
\caption{\label{fig:corner} The posteriors for the amplitude and
  exponent of the SGWB power spectrum, as defined in \eq{e:OmegaOfF}.}
\end{figure}

\emph{Results of the five PTA analysis} -- One important result of
this analysis are the posterior distributions of the SGWB
power-spectrum amplitude $A_{gw}$ and exponent $\gamma_{gw}$ as
defined in \eq{e:OmegaOfF}.  These are shown in
Fig.~\ref{fig:corner}. The amplitude posterior is sharply peaked away
from zero, which is consistent with (or evidence for) the presence of
an SGWB.  As expected, the posteriors are very similar for the two
hypotheses.  This is because, if there are $\lesssim 1500$ pulsars
(see~\cite[Fig.~10]{allen2023hellings}), most of the SGWB information is
provided by auto- rather than cross-correlations, and by
construction, CURN includes these autocorrelations.

\begin{table}[h]
  \centering
  \begin{tabular}{llll}
    \multicolumn{1}{c}{Model}& 
    \multicolumn{1}{l}{$\quad \log_{10}A_{gw}$} & 
    \multicolumn{1}{c}{$\gamma_{gw}$}&
    \multicolumn{1}{c}{$\quad \log_{10}\Omega_{gw}$}\\[0.4ex] 
    \hline
    \noalign{\vskip 0.4ex}
    CURN ($\gamma_{gw}$ free)   & $\quad -14.37^{+0.11}_{-0.12} $   & $\quad 3.63^{+0.27}_{-0.26}$ & $\quad -9.00^{+0.08}_{-0.09}$\\[0.6ex]  
    CURN ($\gamma_{gw}$ fixed)  & $\quad-14.69^{+0.04}_{-0.04}   $  &  $\quad 13/3$              & $\quad -8.92^{+0.08}_{-0.08}$\\[0.4ex] 
    \hline
    \noalign{\vskip 0.4ex}
    HD ($\gamma_{gw}$ free)      & $\quad-14.45^{+0.13}_{-0.15} $ & $\quad 3.76^{+0.31}_{-0.29}$  & $\quad -9.05^{+0.09}_{-0.09}$\\[0.6ex] 
    HD ($\gamma_{gw}$ fixed)     & $\quad-14.72^{+0.04}_{-0.04}$  & $\quad 13/3$                & $\quad -8.98^{+0.08}_{-0.09}$\\[0.4ex]
    \hline
  \end{tabular}
  \caption{\label{tab:pos} Median posterior values for $A_{gw}$ and
    $\gamma_{gw}$, their 68\% containment intervals from Fig.~\ref{fig:corner},
    and $\Omega_{gw}(3\,\text{nHz})$.}
\end{table}

These posteriors, summarized in Table~\ref{tab:pos}, are consistent
with the individual PTA results~\cite{EPTA_new23, miles2025meerkat,
  NANOGrav_new, PPTA_new}.  They are more constraining than the
posteriors obtained by the individual PTAs: the uncertainty ellipses
in the bottom-left plot of Fig.~\ref{fig:corner} are smaller (by about
a factor of two) than those obtained by intersecting the individual
PTA posteriors~\cite{chalumeau2024comparing}.  For example, the
$\text{area} \equiv \int \! \dd\gamma_{gw} \int \! \dd\log_{10}(
A_{gw})$ of our 95.5\% containment ellipse for the HD hypothesis is
48\% of that shown in~\cite[Fig.~1, right]{chalumeau2024comparing}.
As explained earlier, this is as expected for an SGWB: joint analysis
yields more evidence and is more constraining than naive intersection
of the separate PTA results.

The PTA community has established basic criteria for claiming
a detection~\cite{IPTAchecklist}.  Among these, one must show that the data is
consistent with the HD signal hypothesis, and very unlikely (at the
$5\sigma$ level) to arise under the CURN hypothesis.  The estimates of
$A_{GW}$ and $\gamma_{GW}$ cannot help with this, since they are
very similar under both hypotheses (see Fig.~\ref{fig:corner}).

As previously described, one way to assess significance is via
frequentist \textit{null-hypothesis testing}~\cite{pernet2016null}
using a \emph{statistic}. This computes the (noise-marginalized
false-alarm) probability $p$ that signal-like data would occur under
the null hypothesis.  We use quadratic statistics of the form $D(z) =
z^{\mathsf T} \bQ z$, where $z$ is the column vector of timing
residuals, and consider three different standard choices of the square
\emph{filter matrix} $\bQ$ (see~\ref{EM:9}
and~\cite[Eq.~(1)]{BestStatistic}).

While the data are fixed, the inferred pulsar and SGWB noise models-and
hence the CURN null-hypothesis distributions-vary across posterior
samples. We therefore adopt a noise-marginalized
approach~\cite{vigeland2018noise, vallisneri2023posterior}, computing
each statistic and its false-alarm probability for every posterior
realization using the corresponding generalized $\chi^2$
distribution~\cite{hazboun2023analytic, RutgerPvalues}. The resulting
$p$-value distributions are shown in SM Sec.~\ref{s:pvalues}.

\begin{table}[h]
  \begin{tabular}{c|ccccc}
    \hline
    ~statistic~ & ~significance~ & mean & median & 68\% range & 95\% range\\
    \hline
    \multirow{2}{*}{OS} 
    & $\,\,\,\sigma$-units:   & {\bf 4.3}   & 4.8   & (4.3, 5.4)   & (3.9, 5.9)  \\
    & $-\log_{10}p$:          &  5.2   & 6.2   & (5.1, 7.4)   & (4.3, 8.7)  \\
    \hline
    \multirow{2}{*}{NP}
    & $\,\,\,\sigma$-units:   & {\bf 3.3}   & 3.8   & (3.2, 4.3)   & (2.7, 4.9)  \\
    & $-\log_{10}p$:         & 3.3    & 4.1   & (3.2, 5.1)   & (2.4, 6.3)  \\
    \hline
    \multirow{2}{*}{NPMV}
    & $\,\,\,\sigma$-units:   & {\bf 3.3}    & 3.7   & (3.1, 4.3)   & (2.7, 4.9)   \\
    & $-\log_{10}p$:         & 3.3     & 4.0   & (3.1, 5.1)   & (2.4, 6.3)   \\
    \hline
  \end{tabular}
  \caption{\label{tab:pvals} Posterior distributions of the
    false-alarm probability $p$ for $\gamma_{gw}$ free. OS is the
    traditional ``optimal'' statistic, NP is the Neyman--Pearson
    statistic, and NPMV is the robust NP statistic.  Ranges are
    centered about the median, and $p$-values are also translated into
    ``equivalent-$\sigma$'' units.}
\end{table}

We summarize the results for all three statistics in
Table~\ref{tab:pvals}.  None has a mean false-alarm probability $p$
small enough to exceed the $5 \sigma$ detection
threshold~\cite{IPTAchecklist}.  However, note that our analysis
yields a traditional OS detection significance of $\approx 4.3\sigma$,
which is higher than the $\approx 3.2\sigma$ significance reported by
NG~\cite{NG3.2sigma}. This suggests that the ``direct data
combination'' method is working as would be expected in the presence
of an SGWB.

Another way to quantify detection confidence~\cite{IPTAchecklist} is
via the \emph{Bayes factor} $\BF = E_\HD/E_\CURN$, the ratio of HD to
CURN model evidences.  Evaluating these (see~\ref{EM:6}) gives
$\ln(\BF)=10.18\pm0.13$ $\implies$ $\BF = 26 \ke 000\pm 3 \ke 000$.
This is two orders-of-magnitude larger than the $\BF \approx 200$
reported by NG~\cite{NANOGrav_new}. For a real Gaussian random
variable, the corresponding
significance~\cite[Eq.~(8)]{vallisneri2023posterior} of $\sqrt{2 \ln
  (\BF)}\,\sigma \approx 4.5\sigma$ is below the threshold for
confident detection.

\begin{figure}[h]
\centering
\includegraphics[width=1.0\linewidth]{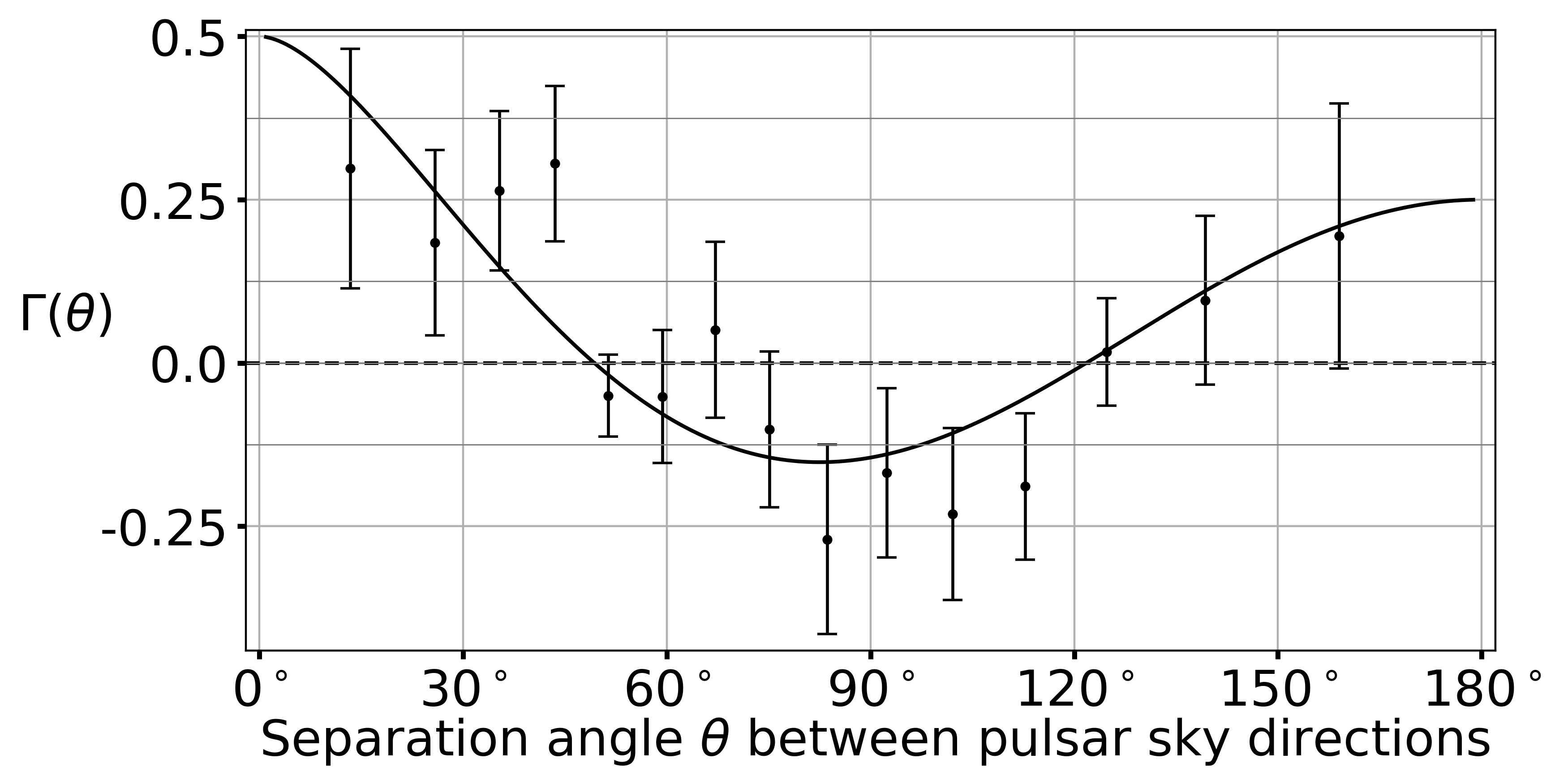}
\vskip -0.1in
\caption{\label{fig:HD} Reconstruction of the mean correlation between
  pulsar pairs as a function of angular separation, compared to the
  Hellings and Downs (HD) prediction~\eq{e:HDmat}. The vertical bars
  indicate the uncertainties in the estimates of the means.}
\end{figure}

We test if the inter-pulsar correlations match those expected from an
SGWB by estimating the angular correlation pattern from the
data~\cite{IPTAchecklist}.  The mean correlation $\Gamma_{ab}$ between
pulsars $a$ and $b$ arising from an SGWB should follow the HD
prediction~\cite{1989NASCP3046...93H, allen2023variance, FAQ}
\begin{equation}
    \label{e:HDmat}
    \Gamma_{ab}  \equiv  \frac{1}{2}(1  +  \delta_{ab})  +  
    \frac{3}{4}(1  -  \cos \theta)
    \! \left[ \ln \Bigl( \frac{1  -  \cos \theta}{2} \Bigr)  -  \frac{1}{6} \right] \! .
\end{equation}
Here, $\theta$ is the angle between unit vectors $\hat n_a$ and $\hat
n_b$ pointing from Earth to pulsars $a$ and $b$, so $\cos \theta =
\hat n_a \cdot \hat n_b$.  This pattern is shown in Fig.~\ref{fig:HD}
as a function of $\theta$.  [The $\delta_{ab}$ in~\eq{e:HDmat} doubles
  the correlation for two pulsars with spatial separation smaller than
  the GW wavelength~\cite{FAQ}.  It plays no role in the
  Fig.~\ref{fig:HD} comparison: the 15 angular bins contain only
  cross-pulsar pairs $a \ne b$, for which the Kronecker-$\delta$
  vanishes.]

An optimal method for ``reconstructing'' the HD correlation from the
data has been derived~\cite{allen2025optimal} but is not yet fully
implemented~\cite{Legendre}.  So, our reconstruction is based on the
same method~\cite{allen2023hellings} previously used by
EPTA~\cite{EPTA_new23} and NG~\cite{NANOGrav_new}, as implemented
in~\cite{nanograv_15yr_repo} (for further details,
see~\ref{EM:8}). Fig.~\ref{fig:HD} shows the result for the \HMP free
$\gamma_{gw}$ CURN posterior samples.  A traditional $\chi^2$
goodness-of-fit test shows that the reconstruction matches the HD
curve well: the reduced $\chi^2$ of $0.74$ is consistent with expected
statistical fluctuations for 15 degrees of freedom

\emph{Conclusion} - We search for an SGWB using public data from five
PTAs.  In effect, we repeat an established analysis, but with a larger
dataset (see~\ref{EM:7}). This is enabled by the physically motivated
step of adopting a single astrophysical/noise model
for each pulsar.

Our results provide the strongest evidence to date that an SGWB
contributes to observed pulsar pulse-arrival-time
fluctuations. However, the statistical significance---assessed using
multiple frequentist and Bayesian methods---remains too low to claim a
detection.

Using existing analysis pipelines enables straightforward comparison
with individual PTA results, with which our findings are consistent,
but it has limitations.  For example, these pipelines assume that for
a stationary process, the Fourier amplitudes in different frequency
bins are statistically independent.  Although this assumption is not
strictly valid~\cite[see Conclusion]{allen2025optimal}, its impact may
be negligible or correctable~\cite{MarcoEtAl}. Likewise, our HD-curve
reconstruction is suboptimal because implementation of an optimal
method~\cite{allen2025optimal} is still in progress~\cite{Legendre}.
Another limitation is the lack of a hierarchical pulsar-noise
model~\cite{RutgerNeedHierarchicalModels,EnsembleNoise}, which could
shift the inferred SGWB amplitude and spectral
slope~\cite{goncharov2025reading}.

For its upcoming DR3 release, the IPTA~\cite{hobbs2010international}
is reanalyzing the five PTA plus CPTA dataset to make new \PAR and
\TIM files. If the primary outcome is adopting common
astrophysical models for the pulsars, then---per the arguments above---we
expect an SGWB search of the DR3 five PTA data to yield results
consistent with ours. But if reanalysis reveals that some data are
incorrect and must be modified or discarded, the results could
differ. Thus, the reliability of our conclusions depends on the
correctness of the five existing PTA analyses.

Notes: (1) Our full analysis code will be released publicly upon
publication. (2) After this work was circulated, a different
data-combination method was proposed~\cite{FrankenStat}. Unlike ours,
that does not use a single consistent astrophysical/noise model for
each pulsar. See also SM Sec.~\ref{s:EFACcuts}.

\emph{Acknowledgments} -- We thank R.~van Haasteren for many useful
conversations, encouragement on this project, and comments on
manuscript drafts.  We also thank S.~Babak, S.~Chen, C.~Clark,
O.~Dodge, B.~Goncharov, M.~Kramer, P.M.~Meyers, M.A.~Papa,
J.D.~Romano, and M.~Vallisneri for helpful feedback.

\bibliography{references}

\begin{thebibliography}{67}
\bibitem [{gss()}]{gss_discovery}%
  \BibitemOpen
  \href@noop {} {}\bibinfo {note} {Our code open to compute the evidence $E$
  will become a pull request in {\texttt{Discovery}}
  \cite{nanogravdiscovery}.}\BibitemShut {Stop}%
\end{thebibliography}

\appendix
\setcounter{emitem}{0}
\section{End Matter}

\noindent
\refstepcounter{emitem}\label{EM:1}\theemitem:\ The diagonal operation
sets to zero all elements where the row and column have different
pulsars, eliminating the inter-pulsar HD
correlations~\cite{BestStatistic}, hence the ``uncorrelated'' in
CURN. It does \emph{not} remove the dependence upon $A_{gw}$ and
$\gamma_{gw}$, which also enter via the diagonal terms. Thus, CURN
does not correspond to the \emph{absence} of GWs, but rather to
``GWs'' which create arrival-time fluctuations \emph{lacking} the
inter-pulsar correlations predicted by GR. Alternatively, CURN
corresponds to an astrophysical or nuclear/QCD physics process of
unknown origin, generating Gaussian rotational fluctuations which are
uncorrelated between pulsars, but have the exact same power-law
spectrum for all pulsars~\cite{FAQ}.  CURN is a more ``conservative''
null hypothesis than that of ``no GWs''.\\

\noindent
\refstepcounter{emitem}\label{EM:2}\theemitem:\ Some previous work
determined $p$ values using ``sky scrambling'' and ``phase shifting'',
which were intended to ``remove any signal from the data''.  The first
leaves the data invariant but assigns a large number of random sky
positions to the different pulsars, thus destroying the HD
correlation~\cite{cornish2016towards, taylor2017all}. The second
retains the pulsar's positions, but randomizes the phases of the
(frequency-domain Fourier amplitudes of the)
data~\cite{taylor2017all}.  However, neither of these methods properly
reproduces the null distribution of $D$~\cite{RutgerPvalues}. For
example, because phase shifting does not modify the squared modulus of
the Fourier amplitudes, it introduces less variance in $D$ than would
arise under the null hypothesis.\\

\noindent
\refstepcounter{emitem}\label{EM:3}\theemitem:\ The data are merged as
follows (also see SM Sec.~\ref{s:workflow}).  If not already present,
rotation counts are added to the \TIM files, which contain pulse TOAs
and uncertainties.  The {\PAR} files, containing astrophysical models
and detector-specific parameters, are modified one pulsar at a time
(the first step of ``manual'' data combination), as follows.
\begin{enumerate}
\item
  For pulsars observed by more than one PTA, a ``reference PTA'' is
  selected via the arbitrary ordering $\rm{NG} > \rm{EPTA} > \rm{PPTA}
  > \rm{MPTA}$. InPTA is not listed because its pulsars are observed
  by other PTAs. Other PTAs that have data for this pulsar
  are called ``target PTAs''.
\item
  Since some pulsars have both J- and B-style names, the name is
  copied from the reference~{\PAR} file to the
  target~{\PAR} files.
\item
  If the pulsar's~{\PAR} files use both TCB and TDB time
  conventions, then quantities whose units contain time are converted
  to TDB with \TEM~\cite{Hobbs:2006cd} tools. See SM Sec.~\ref{s:workflow}.
\item
  The reference~{\PAR} file DM model is replaced with a
  constant-in-time DM model: DM0 is initialized to the {\PAR} file DM
  value at time DMEPOCH, and DM1 and DM2 are initialized to zero.
\item
  A nominal astrophysical model $\beta_0$~\cite[Eq.~7.2]{taylor2021nanohertz}
  (sky position, proper motion, spin frequency, spin-down
  parameters, binary orbital parameters, and DM values) is copied from
  the reference~{\PAR} file to the target~{\PAR}
  files.
\end{enumerate}
Detector-specific parameters in the~{\PAR} files, for example ``FDx''
for narrow-band frequency offsets, are unchanged.  One additional
adjustable phase shift parameter (``JUMP'') per target PTA is
introduced to account for pulse profile template shape differences.\\

\noindent
\refstepcounter{emitem}\label{EM:4}\theemitem:\
The WN covariance matrix $\bN$
characterizes the timing residual fluctuations which remain after
subtracting the deterministic timing model determined by $\beta$, and
stationary power-law-spectrum models for the RN and DM
noise~\cite[Eqs.~7.2 and 7.26]{taylor2021nanohertz},
modeled with 30 and 100 frequency bins respectively. Note that the WN
model is nonstationary: for a given pulsar/backend/PTA, the WN
variance differs between observations.

For pulsars with data from multiple PTAs, the WN parameters are
shifted away from their single-PTA analysis values. This is because
the combined data allow better modeling of the low-frequency RN and DM
noise, leading to more precise estimates of the WN parameters.  In SM
Sec.~\ref{s:EFACcuts}, we show that our conclusions are robust: they
are not changed much by dropping TOAs from backends with the largest
$\efac$ corrections.\\

\noindent
\refstepcounter{emitem}\label{EM:5}\theemitem:\
NUTS is the default sampler for {\DISC}~\cite{nanogravdiscovery},
which provides the log-likelihood gradients needed by Hamiltonian
samplers. It generates samples that are much more independent than the
parallel tempering MCMC sampler~\cite{justinellis20171037579}
previously used by the individual PTAs, and is also much faster.\\

\noindent
\refstepcounter{emitem}\label{EM:9}\theemitem:\ 
The optimal statistic (OS) selects $\bQ$ to maximize the
signal-to-noise ratio
(SNR)~\cite{vigeland2018noise,RutgerPvalues,BestStatistic}. The
Neyman--Pearson (NP) statistic selects $\bQ$ to maximize the detection
probability at fixed false-alarm probability
$p$~\cite{BestStatistic}. The NP minimum-variance (NPMV)
statistic~\cite{BestStatistic} adopts the NP matrix but sets all
diagonal entries of $\bQ$ to zero, eliminating autocorrelation
contributions, making it less sensitive to errors in modeling pulsar
noise~\cite{BestStatistic}.\\

\noindent
\refstepcounter{emitem}\label{EM:6}\theemitem:\ Here, the
\emph{evidence} $E$ is the likelihood marginalized over the
prior~\cite{Jeffreys1961}; $E$ has units $({\rm time})^{-N}$ where $N=
1 \ke 090 \ke 206$ is the number of TOAs (see SM Sec.~\ref{s:EFACcuts}
Table~\ref{tab:efac_cuts}).  Evaluating the evidence integrals is
challenging because the parameter space has $4 \times 121 + 2 = 486$
dimensions for $\gamma_{gw}$ free ($485$ for $\gamma_{gw} = 13/3$).
To compute the evidences we use generalized stepping-stone
sampling~\cite{zahraoui2025generalized, gss_enterprise,
  gss_discovery}, which broadens the sharply peaked likelihood using
$K = 8$ chains ($K-1=7$ steps) while restricting the integration
region. This gives $\ln (\rm{s}^N  \, E_\HD ) = 12 \ke 500 \ke 504.39
\pm 0.10$ and $\ln (\rm{s}^N \, E_\CURN) = 12 \ke 500 \ke 494.21 \pm
0.09$.\\

\noindent
\refstepcounter{emitem}\label{EM:8}\theemitem:\
Our HD reconstruction uses the NG~\cite{NANOGrav_new} choice of 15
angular bins.  The $121 \times 120/2 = 7 \ke 260$ pulsar pairs are
divided equally: each bin has 484 pairs.  For each posterior sample,
the correlations of the pulsar pairs are weighted as
in~\cite{nanograv_15yr_repo}, giving \HMP mean correlations $\mu$ and
their variances $\sigma^2_\mu$ per bin.  For each bin, we then
construct and average the \HMP corresponding Gaussian probability
distributions.  The resulting 15 (posterior estimators of the)
probability distributions of the correlation are sharply peaked.
Fig.~\ref{fig:HD} shows their means (close to medians) and standard
deviations.\\

\noindent
\refstepcounter{emitem}\label{EM:7}\theemitem:\ We deliberately kept
this analysis straightforward and did not ``iterate'' it. This departs
from standard pulsar-astronomy practice, where models (e.g., for
pulsar timing) are iteratively revised. Here, we avoid iterative
refinement and tuning of the full PTA search pipeline. Past experience
with weak-signal detection in LIGO, as well as theoretical studies for
PTAs~\cite{RutgerWarning}, suggests that such iteration can be
misleading.

Because they are not public, we could not include CPTA
data~\cite{CPTA_new} in our analysis. These are from the
Five-hundred-meter Aperture Spherical Telescope (FAST) in Guizhou,
China, which is the world's most sensitive radio telescope. Including
these high-quality FAST data in future analyses would be interesting
and informative.\\


\clearpage
\onecolumngrid
\setlength{\oddsidemargin}{0.5cm}  
\setlength{\evensidemargin}{0.5cm} 
\setlength{\topmargin}{-0.5cm}      
\setlength{\textwidth}{15cm}         
\setlength{\textheight}{22cm}         
\setcounter{page}{100}
\renewcommand{\thepage}{SUP-\the\numexpr\value{page}-99\relax}
\fontsize{12pt}{14pt}\selectfont

\setcounter{secnumdepth}{2}

\section*{Supplementary Material}

Here, we provide additional plots, figures, and tables as supporting material.

\subsection{Data Analysis Workflow}
\label{s:workflow}

\begin{figure}[h]
\centering
\includegraphics[width=1.0\textwidth]{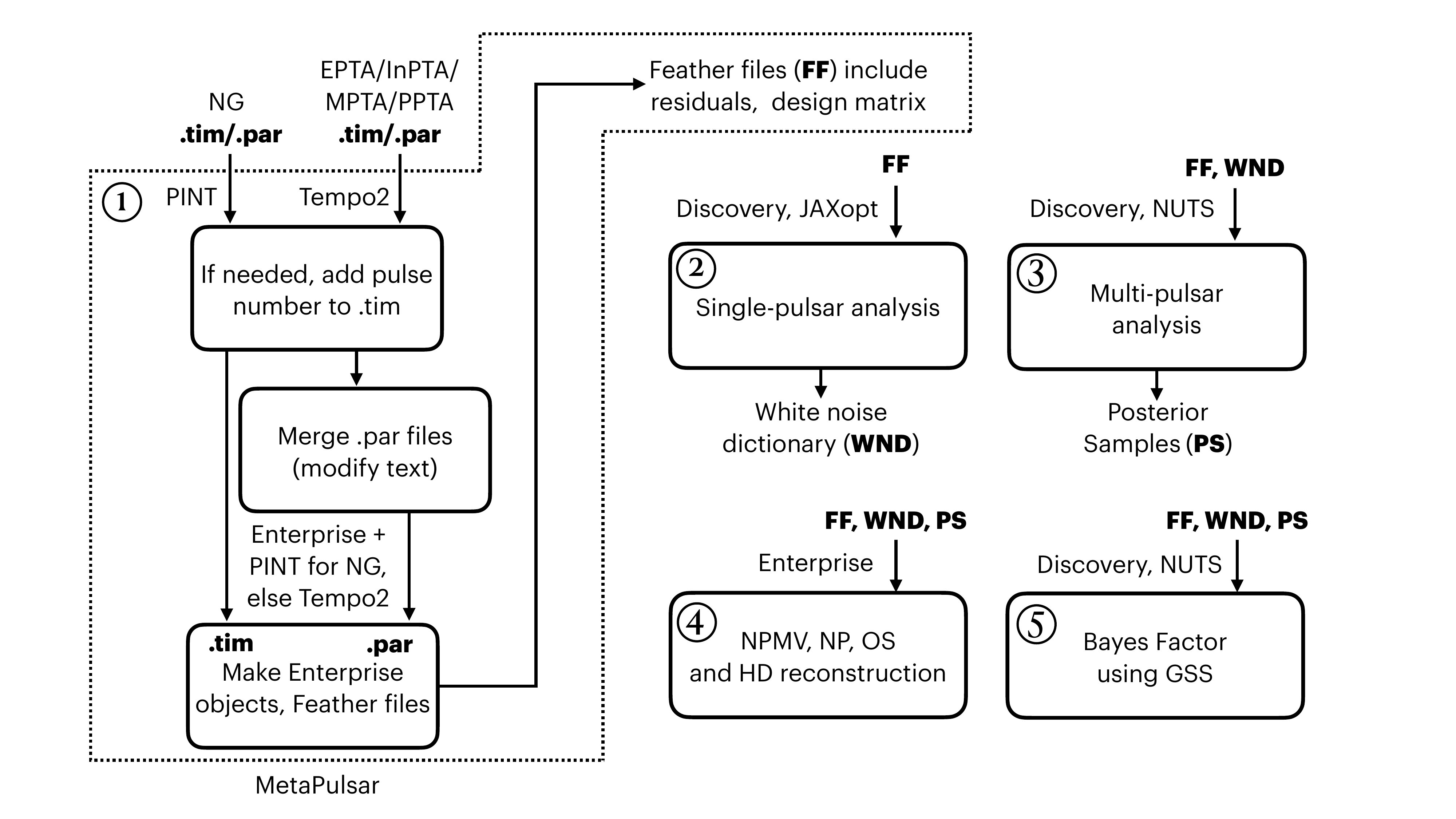}
\vskip -0.1in
\caption{\label{fig:workFlow} A graphical illustration of the data
  analysis workflow. The input data are \PAR and \TIM files from the
  five PTAs.  From these, we produce feather files ({\bf FF}), a white
  noise dictionary ({\bf WND}), and posterior samples ({\bf PS}) which
  are used for further analysis. The circled numbers one to five order
  the analysis steps.}
\end{figure}

Our data analysis workflow is illustrated in Fig.~\ref{fig:workFlow}.  It begins at
the top left with \PAR and \TIM files coming from the five PTAs'
public data releases.  The \PAR files contain parameters such as the
pulsar sky position and spin frequency, and the \TIM files contain the
TOAs.

The PTAs employ pulsar timing programs to prepare these files.  These
programs read and write the files, and are used to fit astrophysical
parameters to the data, by minimizing the timing residuals.  NG uses
the \PINT for this purpose, whereas the other four PTAs use \TEM.

The direct data combination method adopts a common astrophysical model
for each pulsar.  In practice, we do this by modifying the \PAR files,
as described in detail in the paper.  To analyze the data for a given
pulsar, we use a single system of units.  For example, if some of the
\PAR files for a given pulsar have TCB time coordinates, and others have
TDB time coordinates, then we convert all of them to TDB.  Quantities
that contain units of time are scaled in value by this conversion.
For example, pulsar rotation frequencies expressed in TDB are larger
than in TCB by about 16 parts per billion.  Alternatively, if all of
the \PAR files use TDB or TCB exclusively, then
time-coordinate-dependent quantities are not modified.

The situation is simpler for the sky position units. Sky positions can
be specified using right ascension (RA) and declination (DEC) or in
ecliptic latitude and longitude, and the angles can be given in
hour-angles, degrees, or radians.  For constructing a common timing
design matrix $\bM$, we need to ensure that only one combination of
these units is employed for a given pulsar.  This is accomplished when the
sky position is copied from the reference to the target \PAR files.

For \PAR files that specify the ELL1H binary model, there is an
additional complication.  In \PINT, if the Shapiro delay parameter
$H3$ is not specified, then it defaults to zero. However, until a
recent revision, this was incompatible with \TEM, which
did not set a default value, but instead read a random value from
uninitialized memory.  Our solution: if there is no $H3$ specified for
the ELL1H model, then we set binary model to ELL1 model, which is
equivalent to ELL1H with $H3=0$.

Note that while \PINT and \TEM internally use slightly different
values for physical constants such as the solar mass, this makes no
difference in our analysis.

The modified \PAR files and the \TIM files (with pulse counts added)
are then passed to \ENT.  This computes the timing residuals and
design matrix $\bM$ for each pulsar, which are stored in a single \FEA
file for each pulsar.  The timing residuals are computed using \PINT
for the NG TOAs, and \TEM for the other TOAs.  These timing residuals
are what remain after subtracting an identical deterministic model for
pulse arrival time from all TOAs of all PTAs.  The \FEA files, one per
pulsar, are the inputs for all remaining analysis, as shown in
Fig.~\ref{fig:workFlow}. The \PAR and \TIM files are no longer needed.

\begin{table}
  \caption{Parameter name equivalences between \PINT and \TEM.}
  \centering
  \begin{tabular}{lll}
    \hline
    Physical meaning & \PINT & \TEM \\
    \hline
    Derivative of projected semi-major axis   $\quad$  & A1DOT  $\quad$ & XDOT \\
    Eccentricity                              $\quad$  & ECC  $\quad$ & E \\
    The first time derivative of eccentricity $\quad$  & EDOT  $\quad$ & ECCDOT \\
    Shapiro delay parameter                   $\quad$  & STIGMA  $\quad$ & STIG, VARSTIGMA
  \end{tabular}
  \label{tab:aliases}
\end{table}

We perform a ``sanity test'', to check that for a given pulsar, the
separate \ENT objects produced for each PTA have identical
astrophysical model parameters.  Although these parameters were
identical in the \PAR files read by \ENT, if the resulting objects are
queried, the names in those produced with \PINT are different than in
those objects produced by \TEM.  The correspondence is shown in
Table~\ref{tab:aliases}.  So, after checking that the parameters have
identical values and equivalent names, the \ENT objects are modified
to recognize all of the names.  This ensures that the
design matrix $\bM$ contains the correct derivatives in any given
column.

\subsection{False-alarm probability \texorpdfstring{$p$-value}{p-value} distributions}
\label{s:pvalues}

Here, we show histograms of the posterior values of the false-alarm
probability for three different detection statistics.

\begin{figure}[h]
  \hspace*{-3mm}
  \includegraphics[width=0.7\textwidth]{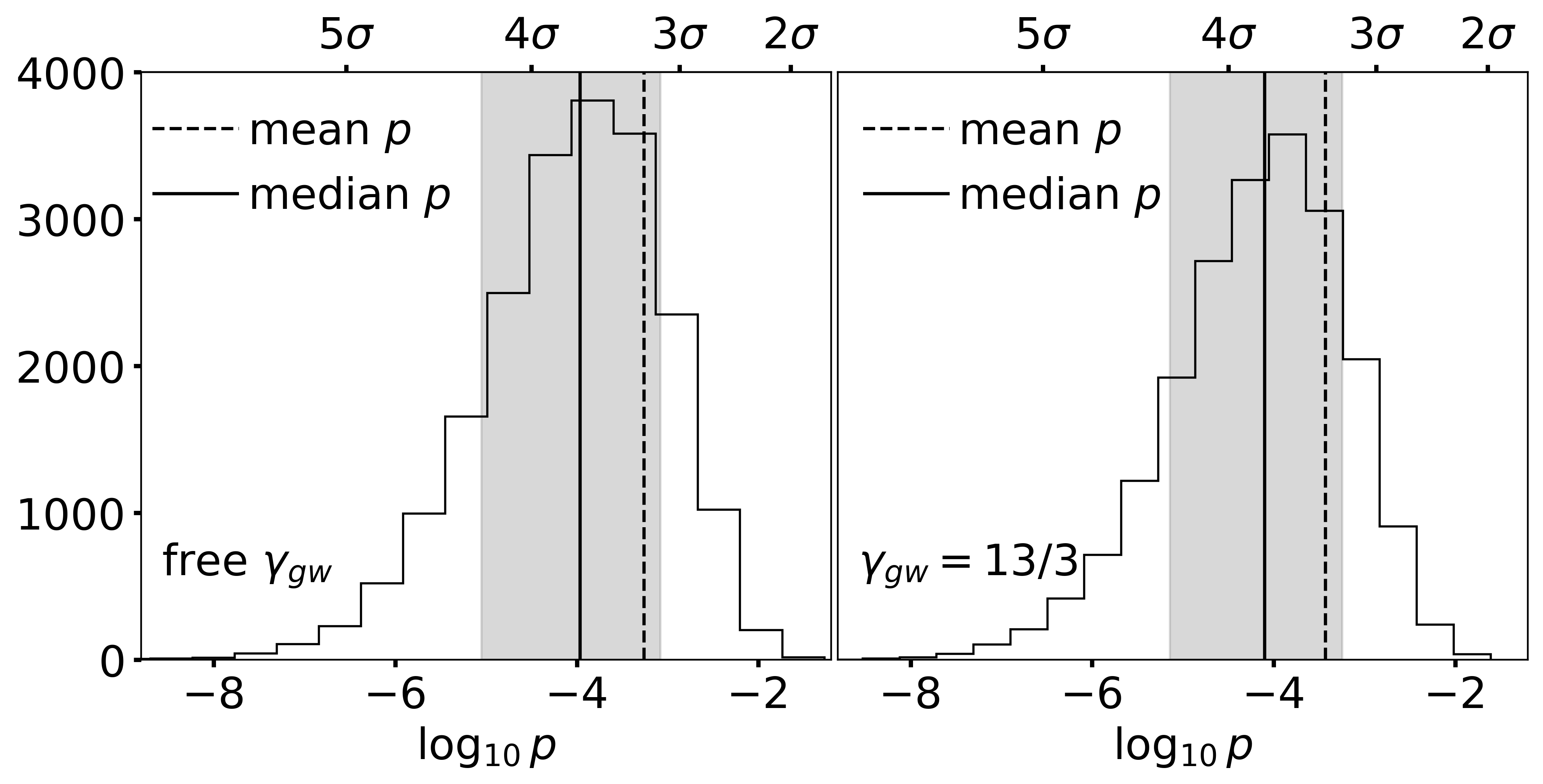}
  \vskip -0.1in
  \caption{Histograms of NPMV detection statistic $p$-values for the
    \HMP CURN-hypothesis posterior samples.  The left panel has
    $\gamma_{gw}$ free, and the right has $\gamma_{gw} = 13/3$. The
    upper horizontal axis is one-sided-Gaussian equivalent
    significance.}
  \label{fig:snr}
\end{figure}

We begin, in Figure~\ref{fig:snr}, with the posterior $p$-value
histograms for the NPMV statistic.  The mean/median false-alarm
probabilities are $p=5.5 \times 10^{-4}$/$p=1.1 \times 10^{-4}$ for
free $\gamma_{gw}$ and $p=3.7 \times 10^{-4}$/$p=8.0 \times 10^{-5}$
for fixed $\gamma_{gw} =13/3$.  Respectively, these correspond to
$3.3\sigma$/$3.7\sigma$ and $3.4\sigma$/$3.8\sigma$ equivalent
Gaussian significances.  The most likely range is shown in
Fig.~\ref{fig:snr} as the band that contains 68\% of the posterior
points about the median. These correspond to significances in the
range from $3.1\sigma$ to $4.3\sigma$ for free $\gamma_{gw}$ and
$3.3\sigma$ to $4.3\sigma$ for $\gamma_{gw}=13/3$.

\begin{figure}[h]
\centering
\includegraphics[width=0.7\linewidth]{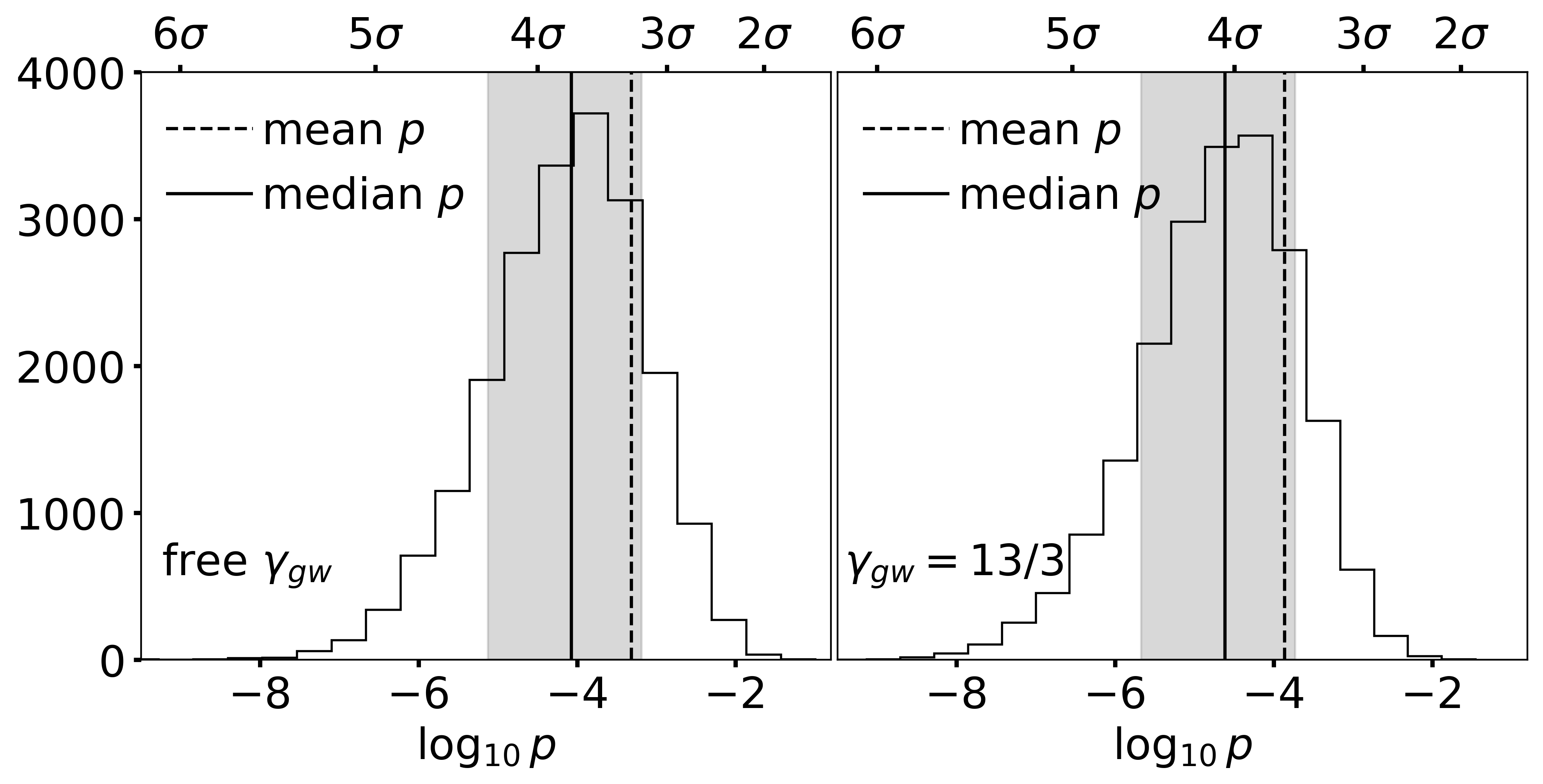}
\vskip -0.1in
\caption{Histograms of NP detection statistic $p$-values for the \HMP
  CURN-hypothesis posterior samples.  The left panel has $\gamma_{gw}$
  free, and the right has $\gamma_{gw} = 13/3$. The upper horizontal
  axis is one-sided-Gaussian equivalent significance.}
    \label{fig:snr_np}
\end{figure}

\begin{figure}[h]
\centering
\includegraphics[width=0.7\linewidth]{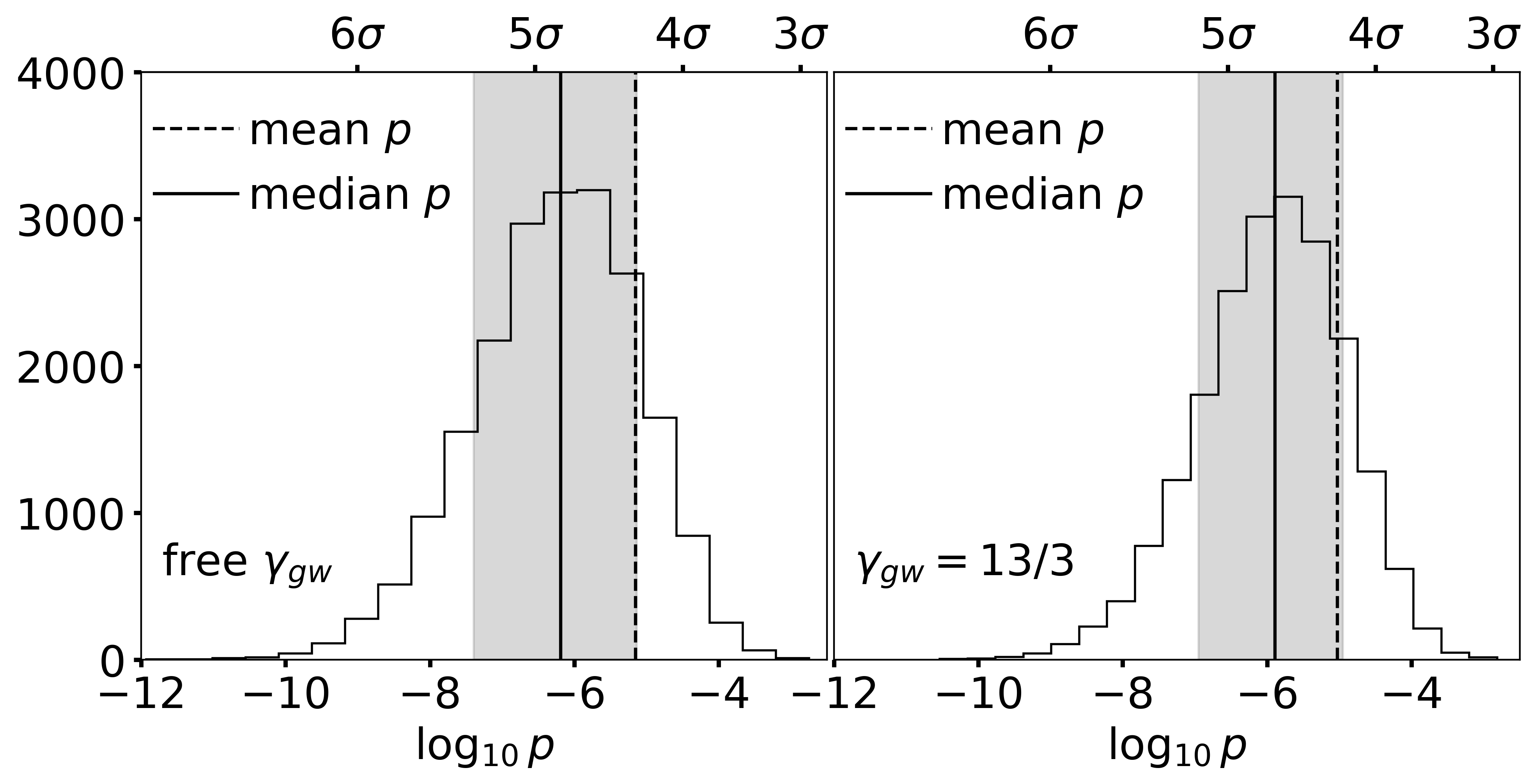}
\vskip -0.1in
\caption{Histograms of the traditional optimal detection
  statistic (OS)
  $p$-values for the \HMP CURN-hypothesis posterior
  samples.  The left panel has $\gamma_{gw}$ free, and the right has
  $\gamma_{gw} = 13/3$. The upper horizontal axis is
  one-sided-Gaussian equivalent significance.}
    \label{fig:snr_dfcc}
\end{figure}

Figures~\ref{fig:snr_np} and~\ref{fig:snr_dfcc}, are the corresponding
histograms for the Neyman--Pearson (NP) statistic and the traditional
``optimal statistic'' (OS).  The means, medians, and widths of these
distributions are reported in Table~\ref{tab:pvals} of the paper.

~

Why do we consider different statistics? Normally, the NP statistic
would be preferred.  However, because its filter matrix $\bQ$ contains
both diagonal (autocorrelation) and off-diagonal (cross-correlation)
terms, it is not robust against errors in modeling pulsar noise. In
contrast, the OS and NPMV statistics use filter matrices that contain
only off-diagonal terms, making them less sensitive to errors in modeling of
the pulsar noise.  Details may be found
in~\cite{BestStatistic}.

\subsection{Effects discarding ``outlier backend'' TOAs}
\label{s:EFACcuts}

\begin{figure}[h]
\centering
\begin{minipage}{0.48\linewidth}
    \centering
    \includegraphics[width=\linewidth]{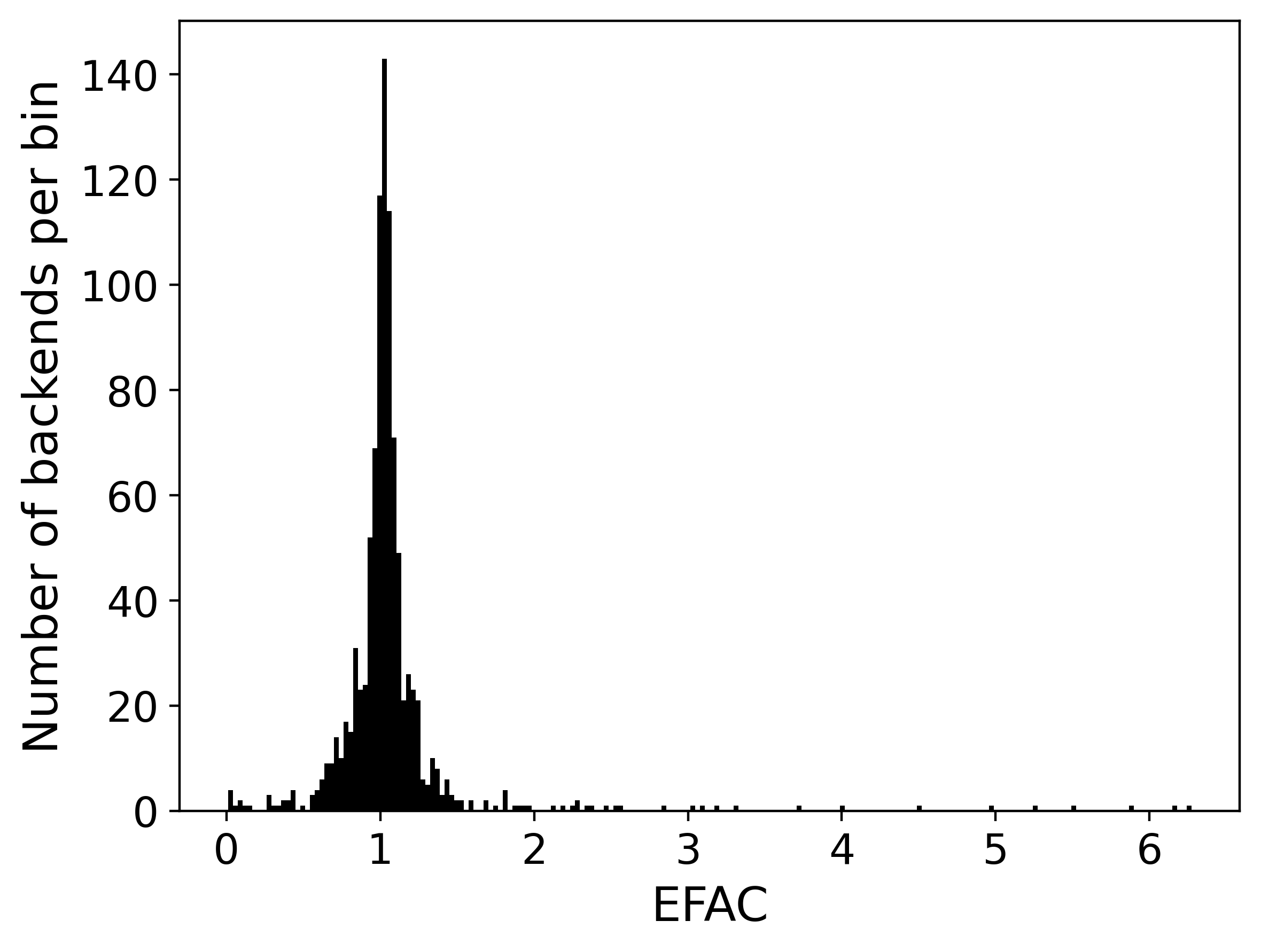}
\end{minipage}
\hfill
\begin{minipage}{0.48\linewidth}
    \centering
    \includegraphics[width=\linewidth]{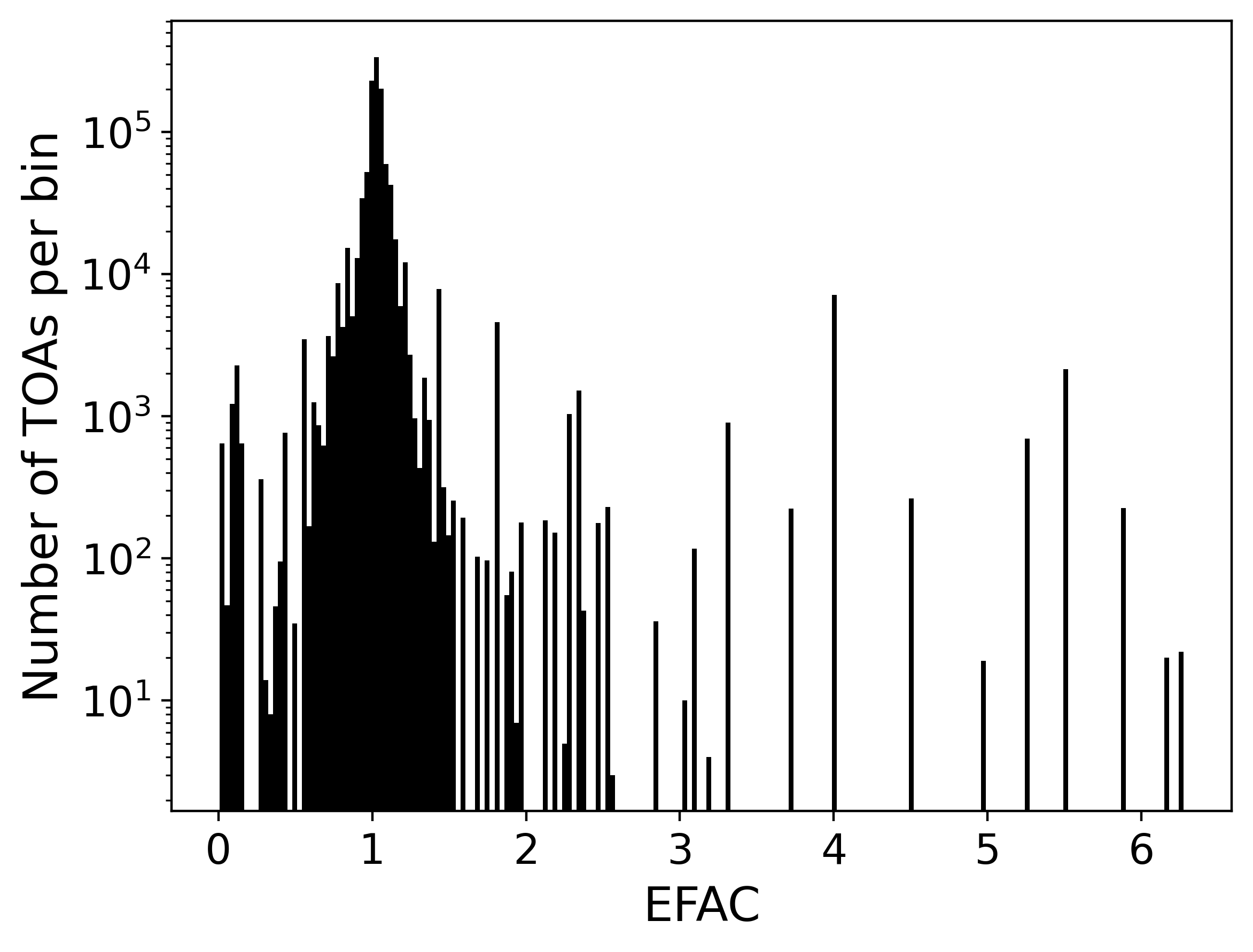}
\end{minipage}
\vskip -0.1in
\caption{
Histograms showing how many backends and TOAs have a given value of $\efac$.}
\label{fsup:n_backends}
\end{figure}

In this work, we analyze the full five PTA dataset, which has $1 \ke
090 \ke 206$ TOAs.  We were curious to see if the results would change
if we discarded ``outlier'' data. By this, we mean data for which the
single pulsar analysis gives WN parameters differing significantly
from those naively implied by the \TIM file contents.  We
flag such outliers via $\efac$, whose nominal value is (close to)
unity, as shown in Fig.~\ref{fsup:n_backends}.

To see the effects of discarding outliers, we drop TOAs from backends
where $\efac$ lies outside the range $[1/Q,Q]$ for $Q=10,4,3$ and $2$.
The number of backends/TOAs and fraction of data eliminated by these
cuts are shown in Table~\ref{tab:efac_cuts}.

\begin{table}[h]
\caption{Summary of EFAC cuts and their impact on backends and TOAs.}
\label{tab:efac_cuts}
\begin{ruledtabular}
\begin{tabular}{lrrrrr}
  EFAC & Backends  & Backends  & TOAs  & TOAs  & \% TOAs  \\
  range &  discarded &  retained &  discarded &  retained &  discarded \\
\hline
  $[0,\infty)$ & 0                               & 976 & 0                   & \mbox{$1 \ke 090 \ke 206$} & 0 \\
  $[0.10,10]$  & 7 PPTA                          & 969 & \mbox{$1 \ke 915$}  & \mbox{$1 \ke 088 \ke 291$} & 0.18 \\
  $[0.25,4]$   & 6 InPTA, 1 NG, \phantom{1}9 PPTA            & 960 & \mbox{$8 \ke 226$}  & \mbox{$1 \ke 081 \ke 980$} & 0.75 \\
  $[0.33,3]$   & 12 InPTA, 2 NG, 12 PPTA         & 950 & \mbox{$17 \ke 002$} & \mbox{$1 \ke 073 \ke 204$} & 1.56 \\
  $[0.50,2]$   & 1 EPTA, 24 InPTA, 4 NG, 18 PPTA & 929 & \mbox{$21 \ke 340$} & \mbox{$1 \ke 068 \ke 866$} & 1.96 \\
\end{tabular}
\end{ruledtabular}
\end{table}

\begin{figure}[h]
\centering
\includegraphics[width=0.50\linewidth]{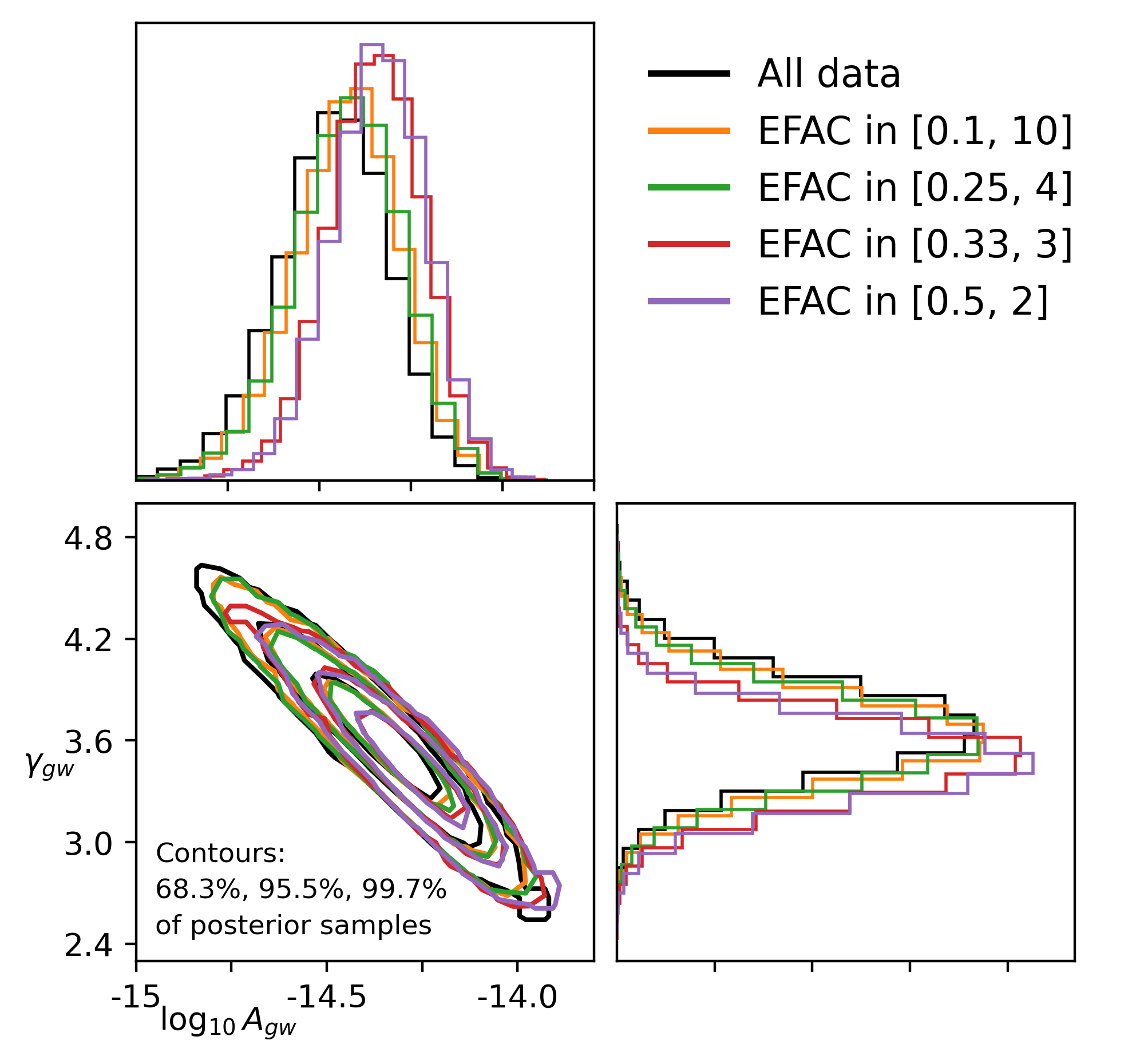}
\vskip -0.1in
\caption{
\label{fsup:thresholds}
  We illustrate the effects of using different $\efac$ thresholds to
  eliminate data from the most poorly modeled backends.  The black
  line shows the CURN posterior from Fig.~\ref{fig:corner} in the
  paper.  The other curves are constructed after dropping TOAs as
  shown in Table~\ref{tab:efac_cuts}, and demonstrate that the CURN
  posteriors are not significantly affected by the data cuts. Note:
  each curve shows $20\ke 480$ posterior samples.}
\end{figure}

The impact of dropping these ``outlier'' TOAs is negligible, as can be
seen from the $A_{gw}$ and $\gamma_{gw}$ posteriors shown in
Fig.~\ref{fsup:thresholds}. These should be compared with
Fig.~\ref{fig:corner} in the paper.  The tighter data cuts do not
significantly affect the posteriors or other conclusions.  We conclude
that our ``all are welcome'' approach to data inclusion has not
significantly skewed or affected our results, but note that discarding
the ``outlier'' data systematically increases the inferred amplitude
of the SGWB.

A recently posted paper~\cite{FrankenStat} comments about ``anomalous
WN parameters'' in a preliminary draft of this work.  These arose
because that earlier analysis contained phase disconnections between
some TOAs observed with different backends.  This was also the reason
that 5 pulsars were given ``special'' treatment there.  We fixed these phase
disconnection issues by by adding rotation counts to the input
files (see \ref{EM:3} and SM Sec.~\ref{s:workflow}).

\end{document}